\PassOptionsToPackage{table}{xcolor}
\documentclass[sigconf]{acmart}
\AtBeginDocument{%
  }

\usepackage{array}
\usepackage{multirow}
\usepackage{dblfloatfix}
\usepackage{enumitem}
\usepackage{fontawesome5}
\setlist[itemize]{leftmargin=1.2em,itemsep=2pt,topsep=2pt,parsep=0pt}
\setlist[enumerate]{leftmargin=1.5em,itemsep=2pt,topsep=2pt}

\setcopyright{acmlicensed}
\copyrightyear{2018}
\acmYear{2018}
\acmDOI{XXXXXXX.XXXXXXX}
\acmConference[Conference acronym 'XX]{Make sure to enter the correct
  conference title from your rights confirmation email}{June 03--05,
  2018}{Woodstock, NY}
\acmISBN{978-1-4503-XXXX-X/2018/06}

\begin{document}

\title{OneRetrieval: Unifying Multi-Branch E-commerce Retrieval with an Editable Generative Model}

\author{Xuxin Zhang$^{*}$, Ben Chen$^{*\dagger}$, Yue Lv, Siyuan Wang, Yupeng Li, Yufei Ma, Zihan Liang, Tong Zhao \\
Ying Yang, Huangyu Dai, Lingtao Mao, Zhipeng Qian, Xinyu Sun, Chenyi Lei, Wenwu Ou, Kun Gai}
\email{\texorpdfstring{\faEnvelope \quad}{}benchen4395@gmail.com}
\affiliation{%
  \institution{Kuaishou Technology}
  \city{Beijing}
  \country{China}}

\thanks{$^{*}$Equal contribution.} 
\thanks{$^{\dagger}$Corresponding author.}

\renewcommand{\shortauthors}{Chen et al.}

\begin{abstract}
Industrial e-commerce search serves hundreds of millions of items through a multi-branch retrieval stage whose sparse, dense, and collaborative branches carry their respective structural weaknesses and are fused by hand-tuned merging with little joint optimization. Generative retrieval (GR) raises the prospect of collapsing this stage into a single model, yet unification is gated by more than retrieval quality: the inverted-index branch converts below the platform average yet persists as almost the only branch through which operations can inject a newly emerging term within hours without a model update, so a one-model substitute needs to preserve this real-time editability. Most existing GR methods structurally lack it: closed-codebook (SID-based) methods typically bind identifier slots to quantized embeddings fixed at training, while open-vocabulary (string-based) methods largely leave the routing of newly emerging terms to model generalization, generally without an explicit mechanism binding a term to a designated item set.

We present \textbf{OneRetrieval}, a \textbf{one}-model generative \textbf{retrieval} framework built on Keyword-Aligned Encoding (KAE), which ties each identifier position to an interpretable key attribute word rather than a quantized embedding; it pairs competitive recall quality with the real-time editability of the inverted index and is, to our knowledge, the first editable generative method with the potential to take over nearly the entire online retrieval stage. An information-theoretic merging organizes 18 key attribute categories into six codebook groups with density-aware non-uniform capacity; each codebook reserves a small block of slots that operations can bind to newly emerging words after deployment without retraining; and a four-stage supervised fine-tuning pipeline secures retrieval quality and editability jointly.

On a benchmark of five million real-traffic requests, OneRetrieval matches the deep-recall quality of the strongest generative baseline and generally outperforms the remaining dense and generative baselines, with an intervention hit rate over an order of magnitude above that of closed-codebook encodings. Online, replacing the inverted-index branch alone significantly lifts order volume, and extending the replacement to nearly the entire retrieval stage leaves conversion essentially unchanged while significantly improving click-through rate, pointing toward a retrieval stage served almost entirely by one generative model. The system is deployed in out-of-mall search at Kuaishou, serving millions of users and generating hundreds of millions of PVs daily. Our code is available at https://github.com/xuxinzhang/oneretrieval.
\end{abstract}

\ccsdesc[500]{Information systems~Retrieval models and ranking}

\keywords{Generative Retrieval, Unified Retrieval, Keyword-Aligned Encoding, E-Commerce Search, Real-Time Intervention}

\maketitle

\begin{table*}[!t]
\centering
\caption{Comparison of retrieval families by identifier construction and inference-time editability. The final column marks whether operations can register a previously unseen item, brand, or category word and serve it without retraining the model.}
\label{tab:rw}
\setlength{\tabcolsep}{6pt}
\footnotesize
\begin{tabular}{l l l l c}
\toprule
\textbf{Family} & \textbf{Representative methods} & \textbf{Identifier} &
\textbf{Semantic Anchor} & \textbf{Editable for new terms} \\
\midrule
Sparse, inverted index    & BM25, TF-IDF, query expansion & lexical term   & surface string         & \checkmark \\
Model-based sparse        & UniDex                        & quantized semantic ID & learned quantization & $\times$ \\
Dense                     & DPR, ColBERT, SimCSE          & dense vector   & learned similarity     & $\times$ \\
Collaborative             & ItemCF, SASRec, SIM           & item ID        & co-interaction history & $\times$ \\
\midrule
Closed-codebook GR        & DSI, TIGER, LC-Rec, OneSearch & fixed integer code   & quantized embedding    & $\times$ \\
Open-vocabulary GR        & SEAL, GLEN, GenR-PO, GRAM, LTRGR     & free-form span & natural-language text  & $\times$ \\
\textbf{Extensible-codebook GR} & \textbf{OneRetrieval (ours)} &
\textbf{integer code + reserved slot} &
\textbf{key attribute word} & \textbf{\checkmark} \\
\bottomrule
\end{tabular}
\end{table*}

\section{Introduction}
\label{sec:intro}
E-commerce search systems route hundreds of millions of items to hundreds of millions of users, within tens of milliseconds per query. The retrieval stage that feeds the downstream ranker is almost universally implemented as a multi-branch architecture: an inverted-index branch handles lexical recall, dense-vector branches handle semantic recall, and collaborative-filtering branches handle behavioral recall, and the candidate sets are merged and forwarded to pre-ranking and ranking \cite{bm25,dpr,colbert,itemcf,sasrec,sim}. The design is robust and easy to debug, yet each branch carries a structural weakness of its own: the inverted-index branch is lexically faithful but semantically blind, surfacing items that match the query string while missing buyer intent; the dense branches generalize across surface forms but match coarsely and turn over new content slowly; and the collaborative branches inherit a head bias that suppresses long-tail and new-product visibility. Keeping the branches side by side adds system-level costs on top, as the branches are trained independently and fused by a hand-tuned merge that precludes joint optimization, with cross-branch redundancy accepted as the price of robustness (Section~\ref{sec:related}). Generative retrieval (GR), which reformulates retrieval as autoregressive generation of an item identifier conditioned on the query and user context, whether a structured semantic identifier (SID) drawn from a compact codebook or a free-form natural-language fragment of the item text, raises the prospect of collapsing the entire stage into a single jointly optimized model.

\textbf{The Editability Paradox.} Internal measurements show that the inverted-index branch occupies a substantial share of exposure, yet its conversion rate falls well below the platform average. Even so, removing it is hardly an option in production, because it is almost the only branch through which operations can inject a new term, a viral brand, or a marketing slogan into the catalog within hours, without any model update. The branch persists not because it retrieves better, but because it is editable. Replacing it therefore calls for a retriever that is at once stronger and comparably editable.

\textbf{Why Generative Retrieval Has Not Closed the Gap.} Existing GR work splits by identifier choice. Closed-codebook methods, such as DSI\cite{dsi}, TIGER\cite{tiger}, LC-Rec\cite{lcrec}, OneRec\cite{onerec}, and OneSearch\cite{onesearch}, learn compact integer codes via residual or product quantization. Open-vocabulary methods, such as SEAL\cite{seal}, GLEN\cite{glen}, GenR-PO\cite{genrpo}, and GRAM~\cite{gram}, use natural-language fragments. Both families improve retrieval quality over the traditional retrieval, yet editability is rarely addressed, and the omission is structural rather than incidental. Closed-codebook methods bind each slot to a quantized embedding fixed at training time, leaving almost no room to accommodate a newly emerging attribute word at inference. Open-vocabulary methods generate free-form fragments, so whether a query carrying a newly emerging term can be reliably routed to the intended items rests almost entirely on the generalization of the model itself; operations have no explicit mechanism that binds the term to a designated set of items. Beyond the generative families, the discriminative substitution of UniDex~\cite{unidex} likewise fixes its quantized identifiers at training and surrenders the same-day editability of the original branch (Section~\ref{sec:related}). Across these lines, the editability that keeps the inverted-index branch alive remains unrecovered, and the full online retrieval stage can hardly collapse into a single model, even where offline quality favors replacement.

How to retain the real-time intervention capability of the online retrieval stage while securing competitive recall quality within one model is thus the central problem of unified generative retrieval. Here we present \textbf{OneRetrieval}, a \textbf{One}-model generative \textbf{Retrieval} framework that pairs competitive recall quality with the real-time editability of the inverted index. To our knowledge, it is the first editable generative method with the potential to take over nearly the entire online retrieval stage. The framework builds on a Keyword-Aligned Encoding (KAE), which aligns each identifier position with an interpretable key attribute word drawn from a production attribute vocabulary rather than with a quantized embedding fixed at training, so that the language-model prior is exploited at each position and the slot-to-meaning relation stays transparent. Capacity is allocated non-uniformly across positions to match attribute density, and a fixed subset of slots is reserved at training time and bound to newly emerging words only after deployment, so that editability resides in the upstream dictionary, exactly where the inverted index keeps it, rather than in the trained policy. These principles give rise to three contributions:
\begin{itemize} 
\item \textit{An extensible codebook design for real-time intervention.} The KAE codebook follows a four-block layout whose reserved block carries no attribute-word binding during training and is bound to newly emerging words only after deployment, giving operations the same-day editability of the inverted index without any model update. The layout rests on two principled choices: an information-theoretic merging of 18 fine-grained key attribute categories into six groups at the knee of the information-loss curve, and a density-aware non-uniform allocation of per-position capacity. The codebook is globally shared and non-hierarchical by design, a choice supported by a negative ablation on entity-conditional encoding.
\item \textit{Attribute-anchored four-stage supervised fine-tuning.} The pipeline opens with a Stage~0 attribute-SID alignment task, which anchors each populated codebook slot to a concrete attribute word before any retrieval objective is introduced, and closes with a reserved-slot self-routing supervision in Stage~3, which keeps unbound reserved slots emittable at decoding time. Retrieval quality and real-time editability are thus carried by near-disjoint stages and secured jointly rather than traded against each other.
\item \textit{Offline and online evidence toward one-model retrieval.} On a large-scale industrial benchmark and in production A/B tests, OneRetrieval matches the deep-recall quality of the strongest generative baseline while recovering an editability that closed-codebook methods can hardly exhibit by construction. Replacing the inverted-index branch alone yields statistically significant conversion gains, and a second deployment that further replaces the dense branch, leaves conversion without significant change while significantly improving click-through rate. Taken together, the offline and online results demonstrate that a single generative model can largely subsume the entire multi-branch retrievals.
\end{itemize}

\section{Related Work}
\label{sec:related}

We organize the related work around the property that gates the one-model unification pursued in this paper: \emph{inference-time editability}, by which we mean whether a retrieval method can absorb a newly emerging item, brand, or category word (e.g., trending IPs such as LABUBU or POP MART) and route it deterministically to a target item population without retraining. Among classical families, methods built on an inverted-index substrate are almost the only ones to offer this property in production; dense retrieval, collaborative filtering, and existing generative retrieval methods generally do not. OneRetrieval is, to our knowledge, the first generative method to recover it. Table~\ref{tab:rw} summarizes where each family stands on identifier construction and editability.

\subsection{Multi-Branch Retrieval in Industrial Search}

The retrieval stage of an industrial e-commerce search system is almost universally a multi-branch architecture over three families of branches. \emph{Sparse} lexical retrieval, of which BM25~\cite{bm25} and the TF-IDF inverted index are canonical, scores documents by term-level statistics. Its editability resides not in the index structure but in the upstream resources that feed it: an operations engineer binds a new term to a target item set through a synonym or tagging dictionary, and an incremental index refresh propagates the binding to serving within hours, with no learned model retrained. Its weaknesses are the lexical-mismatch problem and an inability to model fine-grained intent. \emph{Dense} retrieval, beginning with DPR~\cite{dpr} and extended by ColBERT~\cite{colbert} and contrastive sentence encoders~\cite{simcse}, replaces lexical matching with inner-product similarity in a learned embedding space; it generalizes across surface forms but matches coarsely, turns over new content slowly, and carries substantial training overhead. \emph{Collaborative} retrieval~\cite{itemcf,sasrec,sim} exploits historical co-interaction signal and is effective on head queries and active users, but inherits a head bias that suppresses long-tail and new-product visibility.

Industrial systems blend the three branches and forward the merged candidates to pre-ranking, accepting cross-branch redundancy as the cost of robustness. The paradigm carries three structural limitations: the branches are trained independently and fused by a hand-tuned merge, which precludes joint optimization; the learned branches rely on vector similarity that is hard to interpret; and they suffer from data sparsity that reinforces a head effect. More recently, UniDex~\cite{unidex} replaces the term-based inverted index with a semantic-ID index produced by a dual-tower quantization model, confirming at production scale that the lexical branch is worth rethinking. The reformulation, however, is discriminative rather than generative and stays within the multi-branch layout, and its identifiers are quantized codes fixed at training; the dictionary-level, same-day editability of the original branch is therefore surrendered rather than recovered.

\subsection{Generative Retrieval}
\label{sec:related:gr}

The rise of LLMs has made GR viable: a model autoregressively produces an item identifier conditioned on the query and user context, replacing the multi-branch pipeline with a single sequence-modeling objective. Two identifier conventions dominate the prior literature. \emph{Closed-codebook identifiers} derive integer codes from item embeddings via residual or product quantization, preserving a compact and addressable code space but training the slot--meaning relation only implicitly. \emph{Open-vocabulary identifiers} use natural-language fragments of the item record, exposing the identifier to the language-model prior directly but inflating the candidate space to the full vocabulary. Neither convention readily supports parameter-free real-time editability, and the limitation is structural rather than a matter of implementation. OneRetrieval introduces a third convention, \emph{extensible-codebook identifiers}, in which the codebook is partitioned into a fixed block bound to attribute words at training time and a reserved block bound to new attribute words only after deployment, so that editability becomes a property of the identifier scheme itself rather than a downstream patch. Table~\ref{tab:rw} makes the comparison precise.

\textit{Closed-codebook identifiers.} DSI~\cite{dsi} introduced learned hierarchical IDs for documents; TIGER~\cite{tiger} applied residual quantization to item embeddings for sequential recommendation; LC-Rec~\cite{lcrec} added multi-stage training that integrates collaborative semantics into a pretrained language model. The family is compact and well aligned with the autoregressive paradigm, but every slot is bound to a quantized embedding fixed at training time, so its meaning is opaque and any new attribute or trending term requires retraining both the quantizer and the policy.

\textit{Open-vocabulary identifiers.} SEAL~\cite{seal} generates $n$-gram substrings of titles; GenR-PO~\cite{genrpo} represents items by multi-span identifiers extracted from raw titles; GRAM~\cite{gram} aligns query and product into a shared textual code; LTRGR~\cite{ltrgr} adds a learning-to-rank objective that optimizes directly toward the final ranking. These methods inherit the language-model prior on natural-language fragments, but they inflate the candidate space and rely on textual signals that are noisy in e-commerce, where titles often contain multiple subjects. Crucially, the open vocabulary leaves little room for controllable editing: whether a specific new term reaches the intended items rests almost entirely on the generalization of the model itself, because the routing is mediated by free-form generation rather than by an addressable slot, and operations have no explicit mechanism that binds the term to a designated set of items.

\textit{Industrial generative retrieval systems.} A recent wave of work has carried GR into production~\cite{onerec,onesearch,chen2026onesearch,zheng2025onevision}. OneRec~\cite{onerec} unifies retrieval and ranking within a single generative model for short-video recommendation; OneSearch~\cite{onesearch} extends this to e-commerce search through keyword-enhanced hierarchical quantization, multi-view behavior injection, and a preference-aware reward; OneSug~\cite{onesug} adapts the paradigm to query suggestion, EGA~\cite{ega} to advertising, and UniROM~\cite{unirom} to unified ad ranking. These deployments show that GR is production-ready, but they share a common blind spot: the identifier vocabulary is generally assumed to be fixed at training time, so a brand or attribute word that appears after the training cut-off can hardly be absorbed without a model update. As an \emph{extensible-codebook} method, \textbf{OneRetrieval} is, to our knowledge, the first GR design in which inference-time editability is a structural property of the identifier scheme, achieved without sacrificing the compact, addressable code space of closed-codebook GR.

\subsection{Attribute Extraction and Codebook Construction}

Attribute extraction in e-commerce has been studied for facet discovery and product understanding through sequence-labeling models such as OpenTag~\cite{opentag} and SUOpenTag~\cite{suopentag}; the typed key attribute vocabulary used in this work is produced offline by an internal attribute-extraction pipeline in this line, which yields globally consistent attribute spans. Codebook construction for retrieval draws on product quantization~\cite{pq}, optimized product quantization~\cite{opq}, and structured discrete representations. Prior GR work treats the codebook as a clustering output, with code slots taken as centroids of an embedding partition and no budget reserved for catalog evolution. We instead treat each codebook as a structured allocation that reserves a block of slots, bound to newly trending attribute words only at intervention time, so the codebook supports catalog evolution without retraining. The merging objective behind our position-specific groups (Section~\ref{sec:method:merge}) draws on mutual-information feature selection and structured sparse coding.

\begin{figure*}[t]
\centering
\includegraphics[width=1.15\textwidth, trim=100 30 0 0, clip]{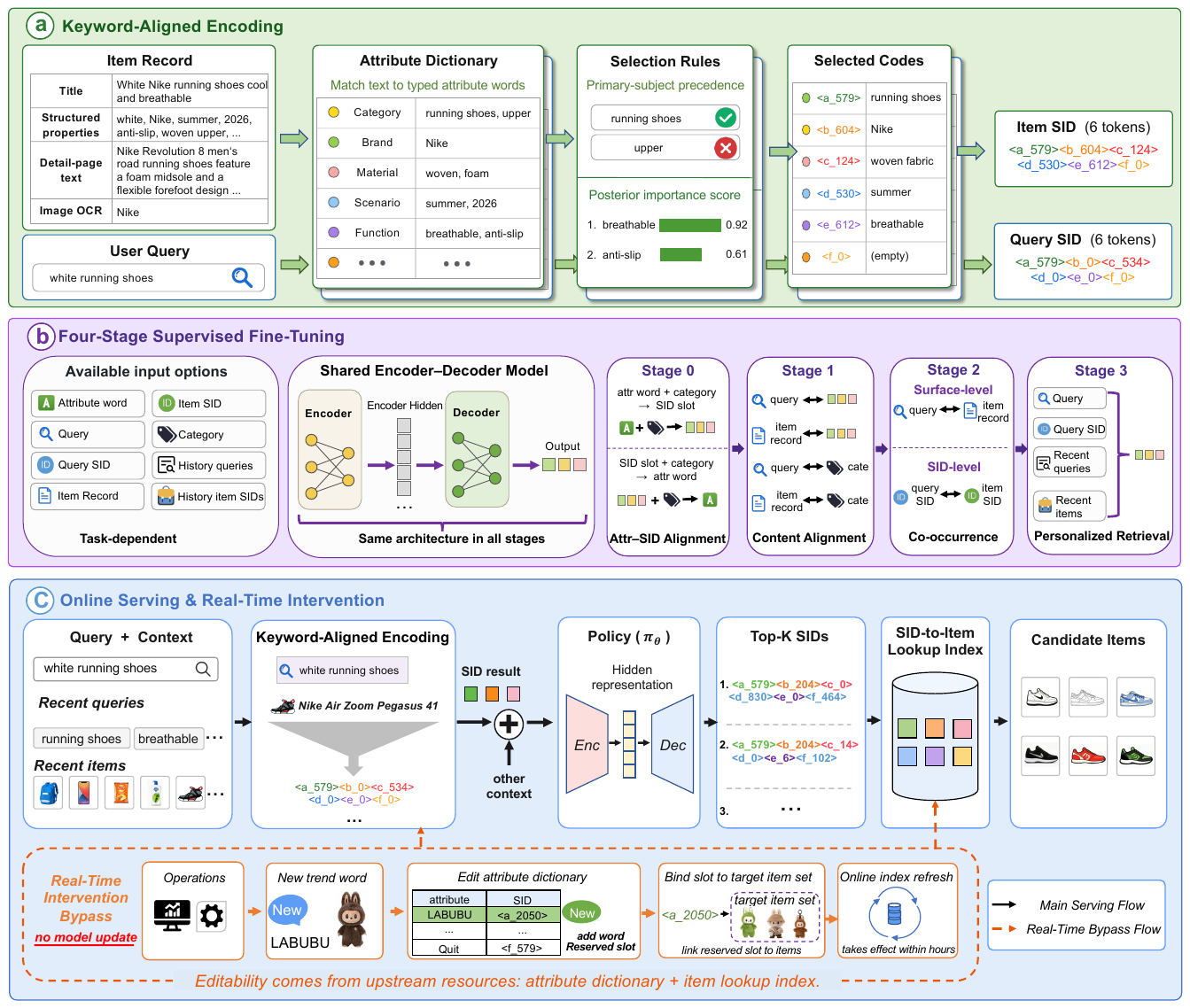}
\caption{Overview of OneRetrieval. (a) KAE maps each item and query to a six-token semantic identifier through a shared attribute dictionary; (b) four-stage supervised fine-tuning anchors the SID alphabet and produces the personalized deployment policy; (c) online serving decodes top-$K$ SIDs by unconstrained beam search, with a dashed reserved-slot bypass that carries the editability of the inverted index into the generative regime without retraining.}
\label{fig:framework}
\end{figure*}

\section{The OneRetrieval Framework}
\label{sec:method}

We formalize retrieval as autoregressive generation over a structured semantic identifier (Section~\ref{sec:method:prob}), outline the offline and online pipelines (Section~\ref{sec:method:overview}), and then present the three coupled designs in pipeline order: attribute-group merging (Section~\ref{sec:method:merge}), the extensible codebook (Section~\ref{sec:method:codebook}), and the SFT pipeline (Section~\ref{sec:method:sft}); Section~\ref{sec:method:hier} explains why the codebook is non-hierarchical.

\subsection{Problem Formulation}
\label{sec:method:prob}
Let $\mathcal{Q}$ denote the space of user queries and $\mathcal{I}$ the catalog of items. Each item $i \in \mathcal{I}$ carries a structured record of a title, structured properties, detail-page text, and image OCR. A user $u$ issues a query $q$ with optional context $\mathbf{c}_u$ comprising the recent search history and recently interacted items. The retrieval stage must return a candidate set $\mathcal{R}(q, \mathbf{c}_u) \subseteq \mathcal{I}$ that maximizes relevance and downstream conversion.

OneRetrieval implements $\mathcal{R}$ as autoregressive generation. Each item $i$ is assigned a structured semantic identifier
\begin{equation}
\mathbf{s}_i = \big(s_i^{(1)}, s_i^{(2)}, \ldots, s_i^{(L)}\big),
\end{equation}
of length $L$, where the $\ell$-th token is drawn from a position-specific codebook $\mathcal{V}_\ell$. At inference time the policy $\pi_\theta$ produces the $K$ most probable SIDs and resolves them into items,
\begin{equation}
\mathcal{R}(q, \mathbf{c}_u) \;=\; \mathcal{T}\!\Big(\operatorname*{TopK}_{\mathbf{s}}\;\pi_\theta\!\left(\mathbf{s} \mid q, \mathbf{c}_u\right)\Big),
\label{eq:retrieval}
\end{equation}
where the top-$K$ operation is approximated by unconstrained beam search and $\mathcal{T}$ is a precomputed SID-to-item lookup index. Because $\mathcal{T}$ is one-to-many---semantically equivalent items can share a SID---the resolved set typically contains more than $K$ items. The policy factorizes autoregressively, $\pi_\theta(\mathbf{s}\mid q,\mathbf{c}_u)=\prod_{\ell=1}^{L}\pi_\theta\!\left(s^{(\ell)}\mid s^{(<\ell)}, q, \mathbf{c}_u\right)$, and is trained by the maximum-likelihood objective summed over positions:
\begin{equation}
\mathcal{L}(\theta) \;=\; -\!\!\sum_{(q,\, \mathbf{c}_u,\, \mathbf{s}) \in \mathcal{D}}\,\sum_{\ell=1}^{L}\,\log\pi_\theta\!\left(s^{(\ell)} \,\big|\, s^{(<\ell)}, q, \mathbf{c}_u\right),
\label{eq:loss}
\end{equation}
where $\mathcal{D}$ is the set of training triples $(q, \mathbf{c}_u, \mathbf{s})$ whose $\mathbf{s}$ is the SID of the item interacted with under $(q, \mathbf{c}_u)$. Equation~\eqref{eq:loss} is the deployment objective; the four fine-tuning stages of Section~\ref{sec:method:sft} reuse the same per-token form over their stage-specific input--target pairs.

This formulation reduces the design space to two coupled questions: (i) how to construct $L$ position-specific codebooks so that the SID is compact, semantically grounded, and editable; and (ii) how to align $\pi_\theta$ with the SID space so that the linguistic prior of the pretrained backbone is fully exploited. We address them in Sections~\ref{sec:method:codebook} and~\ref{sec:method:sft}, respectively.

\subsection{Framework Overview}
\label{sec:method:overview}

Figure~\ref{fig:framework} summarizes OneRetrieval as three panels: (a) keyword-aligned encoding, (b) four-stage supervised fine-tuning, and (c) online serving with a real-time intervention bypass.

\textbf{Keyword-Aligned Encoding (Figure~\ref{fig:framework}a).}
KAE maps each query and item to a SID against a single shared resource: a production \emph{attribute dictionary} $\mathcal{A}$ that lists each typed attribute word with its key attribute category. Input text fields are matched against $\mathcal{A}$ by a deterministic procedure, each matched word is assigned to one of six merged groups (Section~\ref{sec:method:merge}), and when several words fall in one group a single representative is selected through selection rules including primary-subject precedence and a posterior importance score; concatenating the per-group slots yields the SID. The procedure is applied symmetrically to items and queries, and the query-side SID $\mathbf{s}_q$ serves both as inference-time input and as a co-training signal in Stages~2 and~3. The full procedure is detailed in Section~\ref{sec:method:codebook}.

\textbf{Four-stage supervised fine-tuning (Figure~\ref{fig:framework}b).}
We train the SID-consuming model through four stages: Stage~0 anchors every populated codebook slot to its attribute word, Stage~1 aligns query/item with SIDs, Stage~2 introduces collaborative co-occurrence signal, and Stage~3 produces the deployment policy under personalized retrieval supervision (Section~\ref{sec:method:sft}).

\textbf{Online serving (Figure~\ref{fig:framework}c, top main-serving flow).}
At inference time the encoder produces $\mathbf{s}_q$ from the query and context through the same dictionary-driven encoding used offline, so no neural module other than the policy is invoked on the request path. The policy $\pi_\theta$ then decodes a top-$K$ list of SIDs by unconstrained beam search, each resolved into candidate items through $\mathcal{T}$, and the candidate items are forwarded to the downstream pre-ranker.

\textbf{Real-time intervention bypass (Figure~\ref{fig:framework}c, dashed bypass flow).}
A side path, never exercised during policy training, lets operations promote a newly trending attribute word without any model update, mirroring how an inverted-index branch is edited. The operator adds the word to the attribute dictionary with an unused reserved slot of its merged group, so that any later query or item containing the word is encoded into that slot, and binds the reserved slot to the target item set in $\mathcal{T}$. Both edits are incremental and propagate to serving through the existing index refresh within hours, with neither the policy nor the codebook structure touched; as with the inverted index, editability resides in the upstream resources rather than the learned policy. Section~\ref{sec:method:intervention} explains why the policy remains able to emit the correct identifier at decoding time, even though no training example ever paired the reserved slot with this word.

\begin{figure}[t]
\centering
\includegraphics[width=\columnwidth]{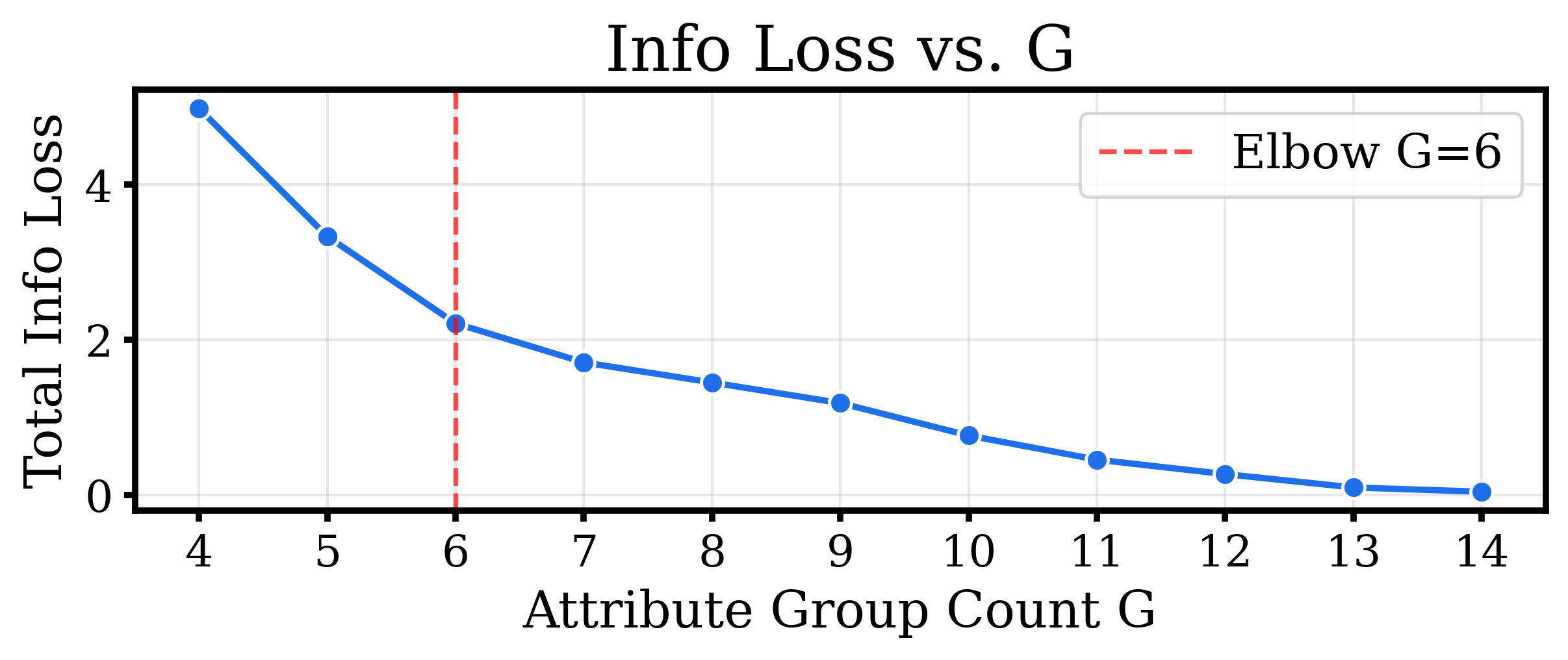}
\caption{Cumulative information loss versus target group count; the second-difference knee falls at six.}
\label{fig:info_loss}
\end{figure}

\subsection{Information-Theoretic Attribute Group Merging}
\label{sec:method:merge}
An internal attribute-extraction pipeline tags substrings of items and queries with one of 18 fine-grained key attribute types: \textsc{entity}, \textsc{brand}, \textsc{anchor}, \textsc{crowd}, \textsc{color}, \textsc{good\_model}, \textsc{specification}, \textsc{material}, \textsc{scene}, \textsc{location}, \textsc{season}, \textsc{marketing}, \textsc{quality}, \textsc{modifier}, \textsc{function}, \textsc{style}, \textsc{pattern}, and \textsc{new}. After PV-based pruning and deduplication the production vocabulary contains about $1.08\times 10^6$ typed attribute words. Mapping each category to its own position would set $L = 18$, inflating inference cost linearly and diluting per-position density, so we seek the smallest $L$ that preserves the discriminative information of the categories and frame the search as information-theoretic agglomerative clustering.

We treat each category $X$ as a Bernoulli activation variable, with $X(i)=1$ when item $i$ carries at least one attribute word tagged as $X$. Writing $p_X = \Pr[X(i)=1]$, $H(X)$ for the binary entropy, and $\mathrm{MI}(X, Y)$ for the mutual information of a pair~\cite{cover}, we measure the loss of collapsing $X$ and $Y$ onto one position by the symmetric conditional entropy
\begin{equation}
\begin{aligned}
\mathrm{IL}(X, Y) &= \tfrac{1}{2}\big(H(X\mid Y) + H(Y\mid X)\big) \\
&= \tfrac{1}{2}\big(H(X)+H(Y)\big) - \mathrm{MI}(X, Y)
\end{aligned}
\label{eq:il}
\end{equation}
which is exactly the residual per-direction uncertainty incurred when one tag must stand in for the other. It is symmetric, non-negative, and equal to one half of the variation of information, so it behaves as a metric over the categories.

\begin{figure}[t]
\centering
\includegraphics[width=1.25\columnwidth, trim=175 120 60 130, clip]{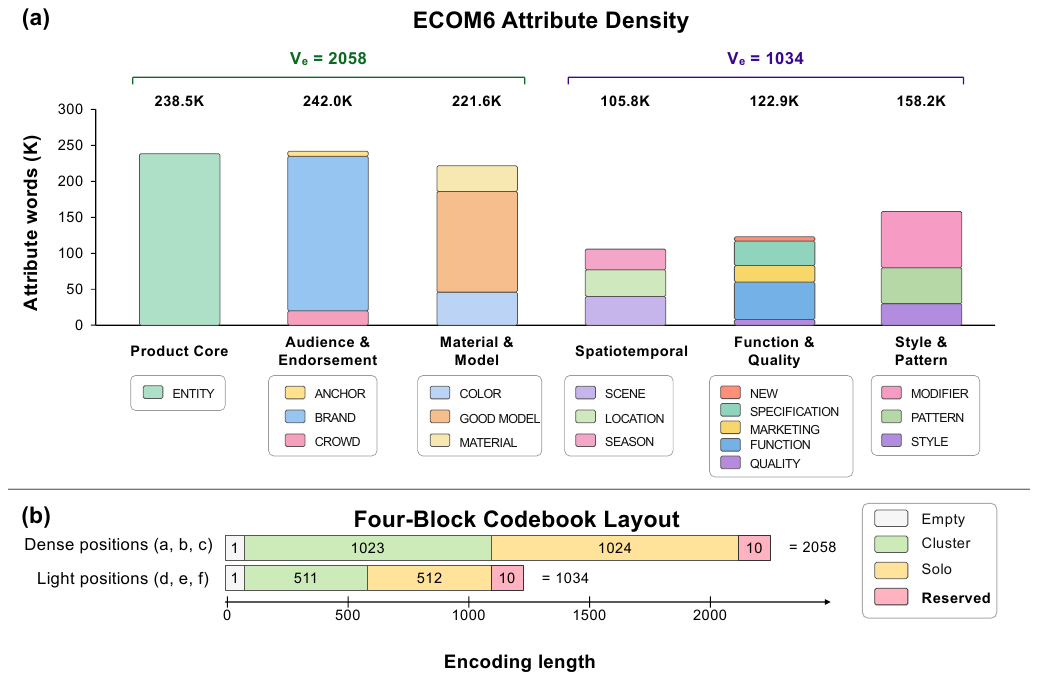}

\caption{Attribute statistics and codebook layout. (a) Attribute-word density of the ECOM6. (b) Four-block layout of each codebook: the empty, cluster, solo, and reserved blocks.}

\label{fig:density}
\end{figure}

We merge the categories by agglomerative clustering under the average cross-group distance $\overline{\mathrm{IL}}(g_a, g_b) = \frac{1}{|g_a||g_b|}\sum_{X\in g_a, Y\in g_b}\mathrm{IL}(X, Y)$, at each step merging the pair with the smallest $\overline{\mathrm{IL}}$. One of the 18 tags is held out: \textsc{entity} is kept as a singleton anchor, since as the noun denoting the bought object it is the primary semantic anchor to which every other attribute attaches. The remaining 17 categories are merged, with the greedy criterion augmented by three lightweight regularizers---within-group entropy and activation-rate homogeneity, and a semantic-coherence prior---and a soft cap on group size, so that the groups stay balanced. To fix the group count, we run the merge to every target count and plot cumulative information loss against it (Figure~\ref{fig:info_loss}): the second difference of the curve peaks at six. We therefore set $G = 6$, and name the partition \emph{ECOM6}. The six groups and their densities appear in Figure~\ref{fig:density}a, whose three densest groups motivate the non-uniform allocation introduced next, and the choice of $L = 6$ is corroborated by the retrieval-quality length scan of Section~\ref{sec:exp:codebook}.

\subsection{Codebook Construction}
\label{sec:method:codebook}

\subsubsection{Non-uniform capacity}
For each merged group $\ell \in \{1, \ldots, L\}$ we construct a codebook $\mathcal{V}_\ell$ of size $V_\ell$, allocated non-uniformly in proportion to the attribute-word density of the group. The recommended configuration sets $V_\ell = 2048$ at the three densest positions and $V_\ell = 1024$ at the remaining three, for $\sum_\ell V_\ell = 9{,}216$ core slots over $L = 6$ positions (Figure~\ref{fig:density}b). We denote a configuration as $\mathtt{L}\langle L\rangle\text{-}\mathtt{D}\langle k\rangle$, where $k$ is the number of positions doubled to $2048$ while the rest stay at $1024$; we write a uniform-$1024$ configuration as $\mathtt{L}\langle L\rangle$ and append $\mathtt{+Hier}$ for an entity-conditional variant (Section~\ref{sec:method:hier}). The uniform baseline is thus $\mathtt{L6}$ and the recommended allocation is $\mathtt{L6\text{-}D3}$, deployed as OneRetrieval. The motivation is direct: per-position attribute-to-slot density is the binding constraint on within-position collision, and uniform allocation forces the densest position to pack an order of magnitude more words per slot than the lightest, which dominates the collision rate.

\subsubsection{Four-block layout}
Within each codebook, the $V_\ell$ core slots occupy three contiguous blocks---empty, cluster, and solo---and a fourth reserved block of $V_\ell^{\text{rsv}}$ slots is appended beyond $V_\ell$ (Figure~\ref{fig:density}b):
\begin{itemize}
\item \textbf{Empty slot.} Index $0$ encodes the absence of any attribute word for this group, or a word outside the production vocabulary.
\item \textbf{Cluster slots.} Indices $\big[1,\; V_\ell - V_\ell^{\text{solo}} - 1\big]$ store cluster centroids from $k$-means~\cite{faiss} over the embeddings of tail attribute words in the group, each covering a neighborhood of synonymous words; representative words are retained for Stage~0 supervision.
\item \textbf{Solo slots.} The next $V_\ell^{\text{solo}}$ indices store the most frequent head words, roughly one per slot, with very close surface variants such as bilingual aliases or common misspellings sharing a slot above a similarity threshold. We use $V_\ell^{\text{solo}} = 1024$ at the dense positions and $512$ at the light ones.
\item \textbf{Reserved slots.} The last $V_\ell^{\text{rsv}} = 10$ indices carry no attribute-word binding at training time and are bound to newly trending words after deployment. They are absent from Stages~0--2, and the policy is exposed to them in Stage~3 only as unbound identity-routing targets (Section~\ref{sec:method:sft}), so they remain emittable at decoding time without being tied to any specific word during training.
\end{itemize}

With $L = 6$ and the recommended allocation, each SID is six tokens. The codebook adds $\sum_\ell V_\ell = 9{,}216$ core tokens and $\sum_\ell V_\ell^{\text{rsv}} = 60$ reserved tokens to the backbone vocabulary, $9{,}276$ in total; the $60$-token reserved budget absorbs a weekly trend cycle on our platform with headroom.

\subsubsection{Item record to SID}
For each item we concatenate title, structured properties, detail-page text, and image OCR, and recover its typed attribute words by matching the text against the production vocabulary with an Aho-Corasick automaton~\cite{aho-corasick}. The vocabulary is bootstrapped offline by running the internal attribute-extraction model over historical corpora, so no neural inference is required at encoding time. The same matching is applied to queries, and each matched word is assigned to its merged group.

When several words fall in one group, a single representative is selected through a precomputed \emph{importance table} that combines two criteria. The first is a primary-subject precedence, constructed once offline by asking an LLM, for every co-occurring pair of same-type words, which is the primary subject---for an item record containing both ``ice cream'' and ``mold'' tagged as \textsc{entity}, the LLM resolves the pair to ``mold'', the object actually being sold. The second is a posterior importance score from behavioral statistics such as PV and CTR, which ranks the remaining candidates once the primary subject is fixed. The representative word is mapped to its codebook slot, and concatenating the per-group slots gives the SID. Figure~\ref{fig:framework}a illustrates the pipeline from dictionary matching through representative selection to the final SID.

\subsubsection{Why reserved slots support real-time intervention}
\label{sec:method:intervention}
Let $w_{\text{new}}$ be a newly trending attribute word, $\ell$ its merged group, and $\langle\ell_v\rangle$ an unused reserved slot. The activation flow is described in Section~\ref{sec:method:overview}; here we make precise why correct retrieval follows without any update to $\pi_\theta$. Three properties underpin this behavior: (P1) and (P3) hold by construction, while (P2) is established by supervision.

\noindent\textbf{(P1) Syntactic reachability.} Under unconstrained beam search the decoder may emit any token of the codebook alphabet at any position. Reserved slots are a fixed subset of this alphabet at training time, so the policy retains the capacity to emit them at decoding time even though no training sample binds one to a specific word. Activating $w_{\text{new}}$ therefore requires no change to decoding, only the two incremental edits of Section~\ref{sec:method:overview}, neither of which retrains $\pi_\theta$ or alters the codebook structure.

\noindent\textbf{(P2) Word-agnostic identity routing.} Reserved slots are absent from Stages~0--2, so the policy carries no bias toward any specific reserved index. In Stage~3 a small block of self-routing supervision of the form $\textsc{prefix}(\langle\ell_v\rangle) \to \textsc{prefix}(\langle\ell_v\rangle)$ teaches the policy to treat any such prefix as an identity router: a query whose SID lands on $\langle\ell_v\rangle$ decodes to an item SID beginning with the same $\langle\ell_v\rangle$. This is independent of the attribute word that $\langle\ell_v\rangle$ eventually represents, since the routing is established by the training data rather than by the binding. The routing is a trained behavior rather than a hard guarantee, and its strength is quantified by the intervention probe of Section~\ref{sec:exp:encoding}.

\noindent\textbf{(P3) Encoder-side determinism.} At query time $w_{\text{new}}$ is recovered by the Aho-Corasick scan over the updated dictionary, assigned to group $\ell$, and mapped to $\langle\ell_v\rangle$ by a deterministic table lookup. The slot a word occupies is fixed by the dictionary rather than selected by the policy, so operations retain full control over the binding, and every query containing $w_{\text{new}}$ presents the same single-position prefix $\langle\ell_v\rangle$---precisely the pattern that the self-routing supervision of (P2) trains the policy to map back onto $\langle\ell_v\rangle$.

Closed-codebook GR baselines such as RQ-VAE satisfy none of (P1)--(P3): their codebooks are derived from training-time quantization, every slot is bound to a fixed centroid, and no position carries a semantic anchor against which a new word could be registered. One might ask whether a quantization codebook could be made editable by appending reserved integer codes, but not in any way separable from KAE: a quantization encoder maps an embedding to its nearest centroid, with no controllable route from a specific new word to a specific reserved code, and constructing such a route requires importing exactly the deterministic dictionary of (P3) and the attribute-anchored, self-routed alphabet of (P2), at which point the scheme is no longer a quantizer. Editability is therefore the joint product of dictionary-based encoding, an attribute-anchored alphabet, and self-routing supervision. The probe in Section~\ref{sec:exp:encoding} confirms this empirically: the RQ-VAE counterpart of OneRetrieval reaches a total IHR$@350$ of only $0.0025$, consistent with incidental code collisions rather than controllable binding.

\subsection{Four-Stage Supervised Fine-Tuning}
\label{sec:method:sft}
Following recent industrial GR work~\cite{onesearch}, we adopt BART-base~\cite{bart} as the encoder-decoder backbone, extend its vocabulary with the $9{,}276$ SID tokens of Section~\ref{sec:method:codebook}, and refine it through four supervised fine-tuning stages trained in sequence, each initializing the next and all sharing the objective of~\eqref{eq:loss}. The task templates appear in Figure~\ref{fig:framework}b, and the per-stage data volumes are reported in Section~\ref{sec:exp:setup}.

\textbf{Stage 0: Attribute--SID alignment.}
Stage~0 teaches $\pi_\theta$ the bidirectional mapping between each populated slot and the attribute word it represents, through a forward template ``\textit{Attribute word is $\langle\textsc{attr}\rangle$, category $\langle\textsc{cate}\rangle$, id is:}'' $\to \langle a_v\rangle$ and its inverse, covering every solo and cluster slot. After Stage~0 each populated slot acquires a position-conditioned distribution over compatible attribute words, and the policy internalizes the bias that a word tagged to group $\ell$ should be emitted at position $\ell$. This positional bias also helps a word later bound to a reserved slot be emitted at its correct position; reserved slots themselves are not part of Stage~0.

\textbf{Stage 1: Content alignment.}
Stage~1 aligns the surface forms of queries and items with their SIDs. Four bidirectional tasks pair queries and item titles with their SIDs, while two category-prediction tasks, $q \to \textsc{cate}_q$ and $\textsc{title}_i \to \textsc{cate}_i$, regularize the SID space toward the platform taxonomy. As the stage that supplies the bulk of query-to-SID and item-to-SID supervision, Stage~1 is the principal source of retrieval quality in the ablation of Section~\ref{sec:exp:sft}.

\textbf{Stage 2: Collaborative co-occurrence.}
Stage~2 introduces collaborative signal at both levels. For each $(q, i)$ pair from click and order logs, two surface-level tasks pair queries with item titles and two SID-level tasks pair $\mathbf{s}_q$ with $\mathbf{s}_i$. The SID-level tasks teach the policy that a query and its converted item occupy compatible regions of the SID space, establishing the query-SID-to-item-SID routing on which the reserved-slot mechanism later relies. As the only ablated stage that exposes the policy to query-side SIDs, Stage~2 is identified as the load-bearing stage for editability in Section~\ref{sec:exp:sft}.

\textbf{Stage 3: Personalized retrieval.}
Stage~3 produces the deployment policy. Each query is augmented with its full user-side context---the current-session query $q$, its derived SID $\mathbf{s}_q$, the recent search queries $\textsc{hist}_q$, and a short sequence of SIDs from recently interacted items $\textsc{hist}_s$---and the main task is
\begin{equation}
(q,\; \mathbf{s}_q,\; \textsc{hist}_q,\; \textsc{hist}_s) \;\to\; \mathbf{s}_i,
\label{eq:stage3}
\end{equation}
where $\mathbf{s}_i$ is the SID of the genuinely interacted item. A sliding-window data refresh keeps the personalization signal aligned with the most recent traffic distribution.

\paragraph{Reserved-slot self-routing supervision.}
Within Stage~3 we add a small auxiliary block that exercises every reserved slot as a word-agnostic identity-routing target, extending to reserved slots the routing that Stage~2 establishes on populated ones. It serves one further purpose: a token that never appears as a target accumulates only negative gradient and is suppressed at decoding time, whereas exercising it as an identity-routing target keeps it emittable. The block is tiny, on the order of $10^{-5}$ of the Stage~3 set, and references no specific attribute word, so a reserved slot is trained purely as an identity router and acquires its meaning only when operations bind a word to it at deployment. Together with the broad routing of Stage~2, this supervision grounds property (P2) of Section~\ref{sec:method:intervention}, leaving the binding as the only operation deferred to deployment.

Because retrieval quality and editability are moved by near-disjoint stage subsets, the leave-one-out ablation of Section~\ref{sec:exp:sft} justifies retaining the full pipeline.

\subsection{Why Not Hierarchical Encoding}
\label{sec:method:hier}
A natural extension is hierarchical encoding, in which the identifier of every non-\textsc{entity} attribute word is made conditional on the entity of the item, so that the brand ``ZARA'' could occupy $\langle b_5\rangle$ under the entity \emph{dress} and $\langle b_{17}\rangle$ under the entity \emph{lipstick}. It is appealing on two counts: the \textsc{entity}-singleton constraint of Section~\ref{sec:method:merge} already treats every other attribute as a facet of the entity, and since most attribute words activate under a narrow set of entities, conditioning could reduce within-position density by up to an order of magnitude.

We tested this with two variants---one making all five non-\textsc{entity} positions conditional on the uniform L6 configuration, and one restricting conditioning to the brand position on the recommended L6-D3---and both underperform their matched non-hierarchical baselines (Section~\ref{sec:exp:codebook}). The failure is systematic, with three configuration-independent causes. First, conditioning turns the slot-to-meaning relation from a function into a one-to-many mapping, diluting the language-model prior that Stage~0 is designed to exploit. Second, because every non-\textsc{entity} attribute word recurs under many entities in production, entity-conditional encoding fragments its training signal and weakens tail-position generalization. Third, an autoregressive decoder commits to the entity at position one, so an error there propagates to every subsequent position; the effect is amplified under the unconstrained decoding adopted in this work, where no prefix-tree correction is available. The retrieval policy is therefore non-hierarchical by design---a deliberate methodological stance, justified by the negative ablation of Section~\ref{sec:exp:codebook}, that other industrial generative retrieval systems can adopt directly.
\section{Experiments}
\label{sec:exp}
The empirical evaluation is organized around five research questions that together cover offline retrieval quality, codebook design, encoding paradigm, training pipeline, and online business outcomes. RQ1 asks how OneRetrieval compares against representative sparse, dense, and generative retrieval baselines on the offline benchmark. RQ2 asks why the recommended configuration is Pareto-optimal across three orthogonal codebook design axes: length, allocation, and conditioning. RQ3 asks why KAE is preferred over embedding-quantization encodings such as RQ-VAE, RQ-kmeans, and RQ-OPQ, evaluated jointly on retrieval quality and real-time intervention capability. RQ4 asks how the four-stage SFT pipeline distributes the two design goals, retrieval quality and editability, across its stages. RQ5 asks whether the offline advantage survives industrial deployment and how far a single generative model can extend across the multi-branch recall stage.

\begin{table*}[!t]
\centering
\caption{Offline retrieval results on the click and order test sets. The deployed configuration OneRetrieval is shaded.}
\label{tab:main}
\setlength{\tabcolsep}{2.5pt}
\footnotesize
\begin{tabular}{l l c c c c c c c c c c c c}
\toprule
& & \multicolumn{6}{c}{\textbf{Order}} & \multicolumn{6}{c}{\textbf{Click}} \\
\cmidrule(lr){3-8}\cmidrule(lr){9-14}
\textbf{Family} & \textbf{Method}
& HR$@10$ & MRR$@10$ & HR$@100$ & MRR$@100$ & HR$@350$ & MRR$@350$
& HR$@10$ & MRR$@10$ & HR$@100$ & MRR$@100$ & HR$@350$ & MRR$@350$ \\
\midrule
\multicolumn{14}{l}{\textit{Traditional retrieval}} \\
& BM25                       & 0.0344 & 0.0133 & 0.1230 & 0.0161 & 0.2215 & 0.0166
                              & 0.0583 & 0.0236 & 0.1798 & 0.0276 & 0.2914 & 0.0282 \\
& docT5query                 & 0.0423 & 0.0165 & 0.1640 & 0.0203 & 0.2926 & 0.0210
                              & 0.0754 & 0.0304 & 0.2314 & 0.0355 & 0.3699 & 0.0363   \\
& DPR                        & 0.0612 & 0.0221 & 0.2605 & 0.0283 & 0.4346 & 0.0293
                              & 0.0956 & 0.0367 & 0.3340 & 0.0445 & 0.5027 & 0.0455   \\																
\midrule
\multicolumn{14}{l}{\textit{Generative retrieval}} \\
& TIGER                      & 0.1253 & 0.0632 & 0.2265 & 0.0673 & 0.2624 & 0.0675 
                             & 0.1393 & 0.0684 & 0.2565 & 0.0730 & 0.2982 & 0.0732   \\							                           
& DSI                        & 0.1427 & 0.0720 & 0.2569 & 0.0764 & 0.2967 & 0.0767
                              & 0.1585 & 0.0800 & 0.2855 & 0.0849 & 0.3306 & 0.0852  \\
& LTRGR                      & 0.1566 & 0.0797 & 0.2837 & 0.0846 & 0.3315 & 0.0849
                              & 0.1768 & 0.0900 & 0.3174 & 0.0954 & 0.3682 & 0.0957  \\
& LC-Rec                     & 0.1667 & 0.0810 & 0.3130 & 0.0865 & 0.3751 & 0.0869
                              & 0.1892 & 0.0915 & 0.3537 & 0.0978 & 0.4164 & 0.0982  \\
& OneSearch                  & 0.2551 & 0.1242 & 0.4766 & 0.1328 & 0.5550 & 0.1333
                              & 0.2909 & 0.1443 & 0.5238& 0.1534 & 0.6007 & 0.1539 \\
\rowcolor{gray!12}
& OneRetrieval                & 0.1846 & 0.0825 & 0.4225 & 0.0868 & 0.5482 & 0.0880 
                              & 0.2034 & 0.1010 & 0.4602 & 0.1060 & 0.6055 & 0.1076 \\
\bottomrule
\end{tabular}
\end{table*}

\subsection{Experimental Setup}
\label{sec:exp:setup}

\textbf{Benchmark.}
The offline benchmark is built from 31 consecutive days of search logs on an industrial e-commerce platform. From the first 30 days we randomly sample $5\!\times\!10^{6}$ user request logs with interactions as the training set, and from day 31 we draw a test set of $29{,}964$ click and $29{,}953$ order samples. After deduplication across the training and test sets together with the recent search history of each user, the query side totals $7{,}629{,}195$ queries, and the item side, comprising every target item together with every item in the historical interaction sequence (clicks and orders), totals $20{,}165{,}617$ items. The typed key attribute words used for codebook construction amount to about $1.08\!\times\!10^{6}$ after deduplication.

\textbf{Baselines.}
We compare against two groups of baselines. The traditional group contains BM25~\cite{bm25}, docT5query~\cite{doct5query}, and DPR~\cite{dpr}, and the generative group contains TIGER~\cite{tiger}, DSI~\cite{dsi}, LTRGR~\cite{ltrgr}, LC-Rec~\cite{lcrec}, and OneSearch~\cite{onesearch}. OneSearch is our prior full-pipeline generative system that unifies recall, pre-ranking, and ranking, with a recall stage built on a keyword-enhanced hierarchical quantization. The closed-codebook recall encoding of OneSearch offers essentially no inference-time editability, so on the intervention axis it is bounded by the closed-codebook paradigms (RQ-OPQ) of Table~\ref{tab:encoding}, which attain near-zero IHR by construction. All generative baselines in Table~\ref{tab:main} share the BART-base backbone, the same training data, and the same materialization of up to five items per SID as OneRetrieval.

\textbf{Metrics.}
Offline we report Hit Rate (HR$@K$) and Mean Reciprocal Rank (MRR$@K$) on both the click and order targets at $K\!\in\!\{10, 100, 350\}$. To quantify real-time intervention capability we introduce two complementary metrics. The Intervention Hit Rate (IHR$@K$) is computed exactly as HR$@K$ except that the target is a fabricated item rather than the genuinely interacted one; it maps the generated SIDs back to items and measures the fraction of samples for which the fabricated target, reachable only through a newly injected code, appears among the top-$K$. The Intervention Activation Rate (IAR$@K$) is the word-level counterpart of IHR$@K$: rather than requiring a single fabricated item to be recovered, it measures the fraction of intervention queries for which the injected attribute word is activated in at least one of the top-$K$ recalled results, which matches the production setting in which an operator binds a word to a target item population rather than to a single item. Because IAR$@K$ is defined for any method that can route an injected word, it serves as the metric for comparing OneRetrieval against the editable inverted-index incumbent (Table~\ref{tab:bm25intv}). The intervention signal is identical in kind for click and order samples, so for both metrics we pool the two and report a single \emph{total} value, with the probe that produces the fabricated items and their injected codes detailed in Section~\ref{sec:exp:encoding}. Online we report Item CTR, Buyer count, and Order volume in out-of-mall search, with statistical significance assessed at the $0.05$ level. To probe the search experience beyond the business metrics, we additionally report a manual side-by-side evaluation along three axes, query-item relevance, item quality, and page good rate, together with a per-industry breakdown of the relative CTR gain.

\textbf{Implementation.}
The backbone is BART-base~\cite{bart}, extended with an SID alphabet whose vocabulary size tracks the codebook configuration. Models are trained on H800 GPUs and served on NVIDIA L20 GPUs, and all offline numbers in the main text use unconstrained beam search with beam size $512$. For each generated SID we materialize up to five items through a precomputed SID-to-item lookup index, and all retrieval metrics are computed at the item level on the resulting list; HR$@350$, for example, is the hit rate over the top-$350$ items obtained after the beam-search SIDs are mapped back to items. All four SFT stages use a learning rate of $1\times10^{-4}$ and a batch size of $512$; Stages~0--2 run for $5$ epochs and Stage~3 for $40$, over roughly $0.8$M, $74.3$M, $10.6$M, and $6.4$M stage-specific examples, respectively. The per-position $k$-means cluster count equals the cluster-block size defined in Section~\ref{sec:method:codebook}.
\subsection{Comparison Against Baselines (RQ1)}
\label{sec:exp:main}
Table~\ref{tab:main} reports the offline retrieval quality of OneRetrieval against the three traditional and five generative baselines.

\paragraph{OneRetrieval Reaches Deep-Recall Parity with the Strongest Generative Baseline.} On retrieval quality alone, OneSearch is the strongest baseline, leading at the shallow and medium cut-offs and on every MRR$@K$. This lead follows from the design goal of OneSearch, which is to unify recall, pre-ranking, and ranking into a single cascade optimized end to end for ranking precision: its closed-codebook hierarchical quantization is tuned to place the most relevant item near the top of the list, exactly the behavior that the shallow cut-offs and MRR reward. OneRetrieval targets the recall stage alone, so it trails OneSearch at the shallow cut-offs yet closes the gap as $K$ grows, until at HR$@350$ the two are effectively tied: $0.5482$ against $0.5550$ on order and $0.6055$ against $0.6007$ on click, the latter favoring OneRetrieval. The two methods stand alone at this depth: the next-best baseline, dense DPR, trails by more than $11$ order HR points, and the strongest remaining generative method, LC-Rec, by roughly $17$. Parity does not reach the top of the list, where the MRR$@350$ of OneRetrieval ($0.0880$ on order) stays near LC-Rec and far below OneSearch ($0.1333$); the contribution of OneRetrieval to retrieval quality is therefore deep-list coverage, the property that matters for a recall branch, rather than the precise top placement for which the cascade of OneSearch is optimized.

\paragraph{The Decisive Axis Is Editability, Not Retrieval Quality.} Because the two leaders are tied at depth, the comparison turns on a second axis that Table~\ref{tab:main} does not measure. OneSearch is a closed-codebook method and therefore carries essentially no real-time intervention capability, on the order of the $0.0021$ total IHR$@350$ of RQ-OPQ in Section~\ref{sec:exp:encoding}, a deficit that closed-codebook generative methods can hardly overcome by construction; OneRetrieval instead recovers the editability of the inverted index. Of the two methods at the deep-recall frontier, OneRetrieval is the one that supplies a capability the other can hardly acquire, and the modest precision gap at the shallow cut-offs is the bounded price paid for it. On the joint objective behind unifying the retrieval stage, deep-recall coverage together with same-day editability, OneRetrieval stands as the more viable choice of the two rather than merely an equally strong one. The configuration evaluated here is L6-D3, justified in Section~\ref{sec:exp:codebook}.

\subsection{Codebook Design Choices (RQ2)}
\label{sec:exp:codebook}

\begin{figure}[!t]
\centering
\includegraphics[width=\columnwidth]{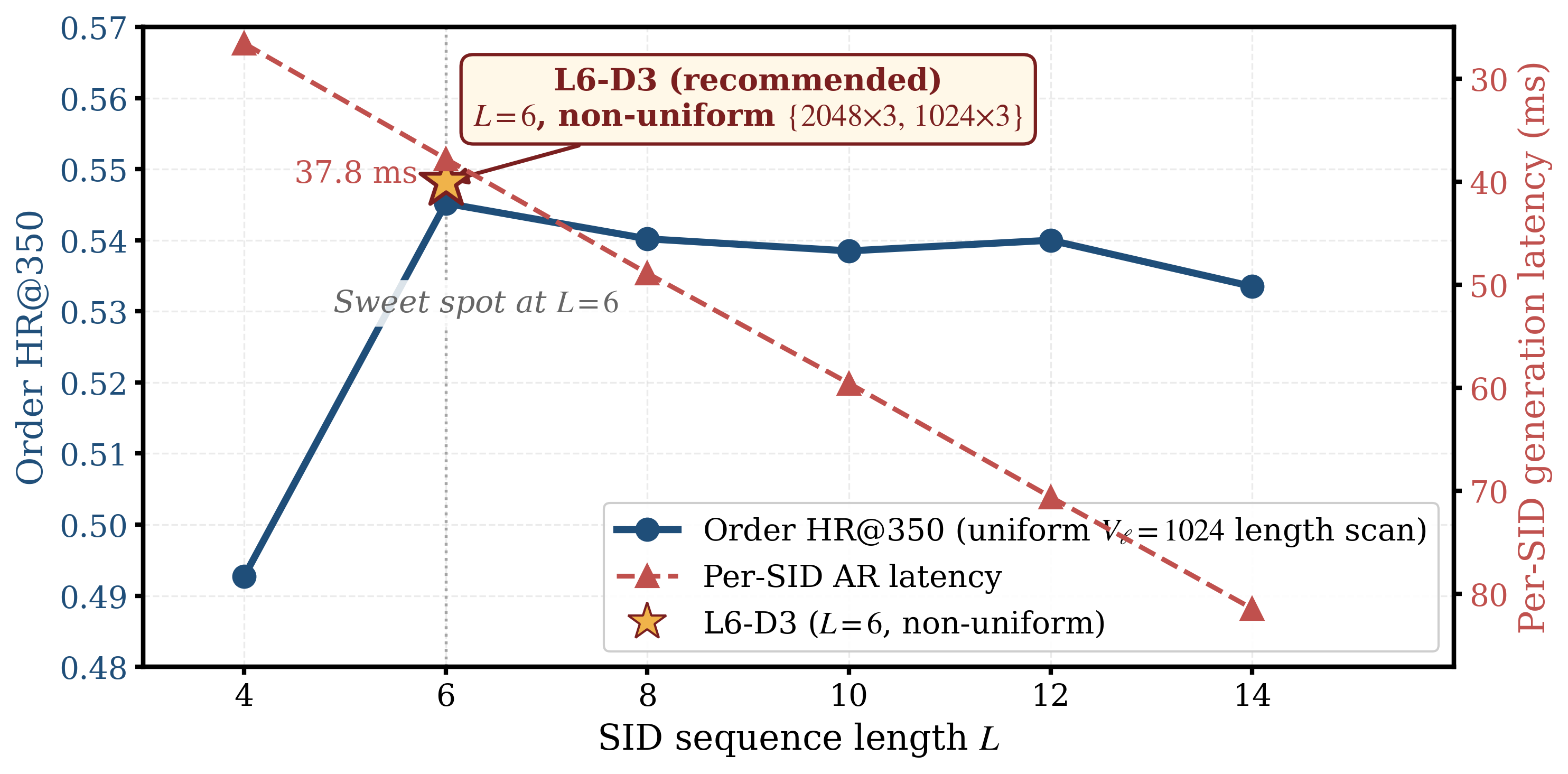}
\caption{Length axis of the codebook design. Order HR$@350$ of the uniform $V_\ell{=}1024$ family peaks at $L{=}6$, while per-SID autoregressive latency grows linearly with $L$; the two curves cross at $L{=}6$. The deployed L6-D3 configuration (gold star) lifts HR above the $L{=}6$ uniform point through density-aware non-uniform allocation.}
\label{fig:codebook}
\end{figure}

The codebook design space decomposes along three orthogonal axes: sequence length $L$, per-position capacity allocation, and conditional versus global encoding. We analyze length first, as it dominates autoregressive inference cost. Figure~\ref{fig:codebook} shows the trade-off within the uniform $V_\ell{=}1024$ non-hierarchical family: Order HR$@350$ peaks at $L\!=\!6$ while per-SID latency grows linearly with $L$, so the two curves cross at $L\!=\!6$, the cost-quality knee. We fix $L\!=\!6$ and analyze the remaining axes at that length.

\paragraph{Allocation Axis: Capacity Tracks Attribute Density.}
Fixing $L\!=\!6$, Table~\ref{tab:allocation} scans five allocations under ECOM6. Order HR$@350$ rises monotonically with total capacity, but what matters is where capacity is placed, not how much. The three leading ECOM6 groups carry far more attribute vocabulary than the trailing three (Figure~\ref{fig:density}a), splitting the positions into a dense head and a light tail, and L6-D3 assigns $V_\ell\!=\!2048$ to exactly the dense head and $V_\ell\!=\!1024$ to the light tail. The intermediate L6-D1 and L6-D2 under-provision the dense head and fall behind despite spending capacity, while L6-D6 edges past L6-D3 only by adding $33\%$ capacity indiscriminately. Because the L6-D1, L6-D2, and L6-D3 differences sit within run-to-run noise, we read L6-D3 as the quality-capacity Pareto knee, the smallest allocation that fully provisions the dense head, rather than the global HR maximum. At $L\!=\!6$ it also surpasses the twice-as-long uniform L12 identifier at half the autoregressive cost, so the non-uniform refinement is captured at the already-optimal length rather than compensating for a suboptimal one.

\begin{table}[!t]
\centering
\caption{Per-position capacity allocation at $L{=}6$ under ECOM6, all non-hierarchical. Total $V$ is the summed core codebook size over the six positions. The deployed configuration L6-D3 is shaded.}
\label{tab:allocation}
\setlength{\tabcolsep}{3pt}
\footnotesize
\begin{tabular}{l l c c c}
\toprule
\textbf{ID} & \textbf{Allocation} & \textbf{Total $V$} & \textbf{Order HR$@350$} & \textbf{Click HR$@350$} \\
\midrule
L6   & $6\!\times\!1024$                       & 6144   & 0.5452 & 0.6033 \\
L6-D1  & $\{2048, 1024\!\times\!5\}$             & 7168   & 0.5459 & 0.6023 \\
L6-D2  & $\{2048\!\times\!2, 1024\!\times\!4\}$  & 8192   & 0.5476 & 0.6035 \\
\rowcolor{gray!12}
L6-D3  & $\{2048\!\times\!3, 1024\!\times\!3\}$  & 9216   & 0.5482 & 0.6055 \\
L6-D6  & $6\!\times\!2048$                       & 12288  & 0.5522 & 0.6072 \\
\bottomrule
\end{tabular}
\end{table}

\begin{table}[!t]
\centering
\caption{Entity-conditional hierarchical encoding at $L{=}6$. Hier.\ scope is the set of positions whose identifier is conditioned on the predicted entity; $\Delta_{HR}$ is the change in Order HR$@350$ against the matched non-hierarchical baseline.}

\label{tab:hier}
\setlength{\tabcolsep}{3pt}
\footnotesize
\begin{tabular}{l l c c c}
\toprule
\textbf{ID} &  \textbf{Hier.\ scope} &
\textbf{Order HR$@350$} & \textbf{Click HR$@350$} & \textbf{$\Delta_{HR}$} \\
\midrule
L6  & none             & 0.5452 & 0.6033 & ref. \\
L6+Hier(all)  & all 5 non-entity & 0.5313 & 0.5872 & $-1.39$ \\
\midrule
L6-D3  & none             & 0.5482 & 0.6055 & ref. \\
L6-D3+Hier(brand) & brand only       & 0.5350 & 0.5916 & $-1.32$ \\
\bottomrule
\end{tabular}
\end{table}

\paragraph{Conditioning Axis: Hierarchical Encoding Underperforms.}
We test the entity-conditional encoding of Section~\ref{sec:method:hier} from two directions, on L6 and on L6-D3: conditioning all five non-entity positions, and conditioning the brand position alone. Both underperform their matched non-hierarchical baselines at equal codebook size (Table~\ref{tab:hier}), losing $1.39$ and $1.32$ Order HR points respectively, the latter despite conditioning only a single position. The loss matches the three configuration-independent mechanisms of Section~\ref{sec:method:hier}, so we retain the globally shared, non-hierarchical codebook.

\paragraph{Synthesis.}
L6-D3 sits on the favorable side of all three axes at once, the cost-quality knee of length, the density-aware optimum of allocation, and the non-hierarchical side of conditioning, and its advantage over twice-as-long uniform identifiers follows only from the three choices reinforcing one another.

\subsection{Encoding Paradigm Comparison (RQ3)}
\label{sec:exp:encoding}

\begin{table}[!t]
\centering
\caption{Comparison of encoding paradigms under a shared L6 codebook layout, training, and evaluation protocol. Order and click HR$@350$ measure retrieval quality; total IHR$@350$ measures real-time intervention capability.}

\label{tab:encoding}
\setlength{\tabcolsep}{4pt}
\footnotesize
\begin{tabular}{l c c c}
\toprule
\textbf{Encoding paradigm} & \textbf{Order HR$@350$} & \textbf{Click HR$@350$} & \textbf{Total IHR$@350$} \\
\midrule
KAE & \textbf{0.5452} & \textbf{0.6033} & \textbf{0.0806} \\
RQ-VAE    & 0.5075 & 0.5516 & 0.0025 \\
RQ-kmeans & 0.5355 & 0.5837 & 0.0030 \\
RQ-OPQ & 0.5376 & 0.5848 & 0.0021 \\
\bottomrule
\end{tabular}
\end{table}

Section~\ref{sec:exp:codebook} fixed the codebook structure of L6-D3; here we isolate the upstream choice of KAE against the embedding-quantization conventions of closed-codebook GR. To attribute any difference to the encoding paradigm rather than to capacity, we fix the uniform $6\times1024$ core layout for all four paradigms (L6; reserved slots are excluded from the $V_\ell$ count but retained in the KAE layout) and vary only how codes are assigned. KAE is compared against three quantization counterparts that encode item embeddings rather than attribute words: RQ-VAE (learned residual quantization), RQ-kmeans (iterative $k$-means residual quantization), and RQ-OPQ (the RQ-kmeans base plus a $2\times64$ OPQ~\cite{opq} code). All four share Stages~1--3, the compute budget, and the evaluation protocol, and differ only in Stage~0, which performs the attribute--SID alignment for KAE but is undefined for the quantization codebooks, whose codes carry no standalone attribute semantics to align against.

To probe intervention offline, we convert the test data into samples that emulate online intervention requests. An LLM (Qwen3-7B) generates $2{,}000$ words outside the attribute dictionary; around each we build a query--item pair and encode it under the paradigm being tested, assigning the word to a randomly chosen reserved slot for KAE and encoding it from its embedding for the quantization baselines. Overwriting the query, item, and codes of $1{,}000$ click and $1{,}000$ order test samples with these fabricated pairs yields $2{,}000$ intervention samples, scored by the total IHR$@K$.

\paragraph{KAE Leads on Retrieval and Far Exceeds Quantizers on Intervention.} KAE attains the highest retrieval quality of the four paradigms on both targets, with Order and Click HR$@350$ of $0.5452$ and $0.6033$, ahead of all three quantizers even without the larger L6-D3 codebook (Table~\ref{tab:encoding}). This retrieval margin is modest; the intervention gap is not. On total IHR$@350$ KAE reaches $0.0806$ against $0.0025$, $0.0030$, and $0.0021$ for RQ-VAE, RQ-kmeans, and RQ-OPQ, more than an order of magnitude higher and a margin no quantizer approaches.  The gap is structural rather than a tuning artifact: as argued in Section~\ref{sec:method:intervention}, a quantization code satisfies none of properties (P1)--(P3), and the small non-zero baseline values reflect incidental code collisions. This comparison therefore establishes only that closed-codebook GR can \emph{hardly} support intervention, not how well OneRetrieval does so in absolute terms; the substantive reference is the incumbent that already possesses editability in production, the BM25 inverted-index branch, against which we compare OneRetrieval on identical injected terms next.

\paragraph{Editability Against the Editable Incumbent.}
The item-level IHR$@K$ above requires a single fabricated item to be fully recovered and is therefore a lower bound on editability, since in production an operator binds an injected word to a target item \emph{population} rather than to one item. The population-level metric is the IAR$@K$ of Section~\ref{sec:exp:setup}: for OneRetrieval a result counts as activated when a decoded SID carries the injected reserved code at its position, and for the BM25 inverted-index branch when a recalled item carries the injected word. The metric is defined for any method that can route an injected word, which makes it the apples-to-apples comparison against the incumbent, whereas the item-level IHR of Table~\ref{tab:encoding} compares only generative paradigms, the quantization codebooks exposing no reserved code to activate. Table~\ref{tab:bm25intv} reports the result: OneRetrieval reaches an IAR$@350$ of $0.553$ against $0.761$ for the inverted index, about three quarters of the incumbent rate, with on average $15.5\%$ of its decoded SIDs carrying the injected code, while more than doubling retrieval quality (Order HR$@350$ $0.5482$ vs.\ $0.2215$, Click $0.6055$ vs.\ $0.2914$). The inverted index remains stronger on raw activation, as expected of a lexical branch whose native operation is surfacing an injected word; the contribution of OneRetrieval is to recover most of that activation inside a single generative model that is far stronger on retrieval and conversion.

\begin{table}[!t]
\centering
\caption{Editability against the editable incumbent, on the same injected terms. Order and click HR$@350$ restate retrieval quality from Table~\ref{tab:main}; total IAR$@350$ is the word-level Intervention Activation Rate.}
\label{tab:bm25intv}
\setlength{\tabcolsep}{4pt}\footnotesize
\begin{tabular}{l c c c}
\toprule
\textbf{Method}& \textbf{Order HR$@350$} & \textbf{Click HR$@350$} & \textbf{Total IAR$@350$}  \\
\midrule
BM25 inverted-index  & 0.2215 & 0.2914 & 0.7610 \\
OneRetrieval     & 0.5482 & 0.6055 & 0.5530 \\
\bottomrule
\end{tabular}
\end{table}
\subsection{Four-Stage SFT Ablation (RQ4)}
\label{sec:exp:sft}

\begin{table}[!t]
\centering
\caption{Leave-one-out ablation of the SFT pipeline under the fixed L6-D3 codebook. Each row removes one of Stages~0--2; Stage~3 is retained throughout, as it produces the personalized deployment policy.}
\label{tab:sft}
\setlength{\tabcolsep}{4pt}
\footnotesize
\begin{tabular}{l c c c}
\toprule
\textbf{Configuration} & \textbf{Order HR$@350$} & \textbf{Click HR$@350$} & \textbf{Total IHR$@350$} \\
\midrule
OneRetrieval          & 0.5482 & 0.6055 & 0.1340 \\
\;\;w/o Stage 0       & 0.5500          & 0.6072          & 0.1020 \\
\;\;w/o Stage 1       & 0.5434          & 0.5967          & 0.1580 \\
\;\;w/o Stage 2       & 0.5485          & 0.6035          & 0.0030 \\
\bottomrule
\end{tabular}
\end{table}

We ablate the pipeline leave-one-out under the fixed L6-D3 codebook, removing one of Stages~0--2 at a time. Stage~3 is retained throughout, since it produces the personalized deployment policy (Section~\ref{sec:method:sft}) and removing it would delete the deployment objective rather than test a swappable component.

\paragraph{Retrieval Quality Is Carried by Stage~1, and Editability by Stage~2 with Stage~0 in Support.} The ablation confirms the per-stage roles assigned in Section~\ref{sec:method:sft} and shows that the two design goals are moved by near-disjoint stage subsets. On retrieval quality, removing Stage~0 or Stage~2 leaves Order and Click HR$@350$ within measurement noise; removing Stage~0 even nudges the point estimate up by a noise-level margin (Order $0.5500$ against $0.5482$), so Stage~0 is retained for the editability and interpretability it provides rather than for retrieval quality. Only removing Stage~1 lowers HR on both targets, and only slightly, consistent with Stage~1 supplying the bulk of the query-to-SID and item-to-SID alignment over which the policy operates. The total IHR$@350$ column tells the opposite story. The full pipeline attains $0.1340$ (above the $0.0806$ of Section~\ref{sec:exp:encoding}, which uses the uniform L6 codebook rather than L6-D3), yet removing Stage~2 collapses it to $0.0030$, while removing Stage~0 lowers it only to $0.1020$ and removing Stage~1 leaves it essentially intact. Stage~2 is the only ablated stage that exposes the policy to query-side SIDs, so it establishes the query-SID-to-item-SID routing that property (P2) of Section~\ref{sec:method:intervention} requires at intervention time, while Stage~0 anchors that alphabet and the positional bias behind it, contributing a smaller but non-trivial share. Because each stage is necessary for the goal it dominates and largely dispensable for the other, no single removal is harmless across both, which justifies retaining the full pipeline.

\subsection{Online A/B Tests (RQ5)}
\label{sec:exp:online}

\begin{table}[!t]
\centering
\caption{Online A/B uplift of OneRetrieval in out-of-mall search, as relative
change of the experiment bucket over the control bucket. The first configuration
replaces the inverted-index branch alone; the second additionally replaces the
dense (vector) branch. $^{\dagger}$ denotes significance at the $0.05$ level.}
\label{tab:online}
\setlength{\tabcolsep}{6pt}
\footnotesize
\begin{tabular}{l c c c}
\toprule
\textbf{Configuration} & \textbf{Item CTR} & \textbf{Buyer} & \textbf{Order} \\
\midrule
Replace inverted-index               & $+0.074\%$           & $+0.450\%$  & $+0.710\%^{\dagger}$ \\
Replace nearly all   & $+0.821\%^{\dagger}$ & $-0.028\%$  & $+0.255\%$ \\
\bottomrule
\end{tabular}
\end{table}

Sections~\ref{sec:exp:main} through~\ref{sec:exp:sft} established the offline Pareto position of L6-D3; we now test whether it survives deployment and how far a single generative model can reach into the multi-branch recall stage. We first replaced the production inverted-index branch with OneRetrieval (L6-D3) in out-of-mall search, under the same unconstrained beam search used offline, with control and experiment buckets each serving $20.0\%$ relative traffic (about $8.2\%$ absolute), balanced over a $7$-day AA window and measured over an $11$-day AB window. The first row of Table~\ref{tab:online} reports the relative change of the experiment over the control bucket.

\paragraph{Single-Branch Replacement Lifts Conversion.}
In out-of-mall search the conversion metrics rise while the click-through rate stays near flat ($+0.074\%$), consistent with OneRetrieval surfacing items of higher conversion propensity rather than inflating clicks. The uplift is attributable to OneRetrieval rather than to retiring a poorly converting branch: a separate $7$-day arm that removed the inverted-index branch without enabling OneRetrieval produced no statistically significant business movement.

\paragraph{Manual Evaluation Confirms a Better Search Experience.}
To verify that the business gains reflect a genuinely better search experience rather than click inflation, we conducted a manual side-by-side evaluation following the protocol of OneSearch~\cite{onesearch}. Trained annotators compared the items that OneRetrieval and the production baseline surfaced at identical exposure positions and labeled each pair as Good, Same, or Bad (GSB) along three axes, with the reported number being the net win rate, the fraction of Good cases minus the fraction of Bad cases. OneRetrieval is classified as a \emph{win} on all three axes: $+0.82\%$ on query-item relevance, $+1.36\%$ on item quality, and $+0.54\%$ on the page good rate that jointly reflects relevance and quality. The gains trace to keyword-aligned encoding, which resolves a query into its attribute words and recalls items matching the underlying intent rather than the surface string, so the recalled set is both more relevant to the query and of higher item quality than the lexically matched set returned by the inverted-index branch. Because the win classification holds uniformly across all three axes, the improvement to the production experience is stable rather than driven by a single outlier dimension.

\begin{figure}[t]
\centering
\includegraphics[width=\columnwidth]{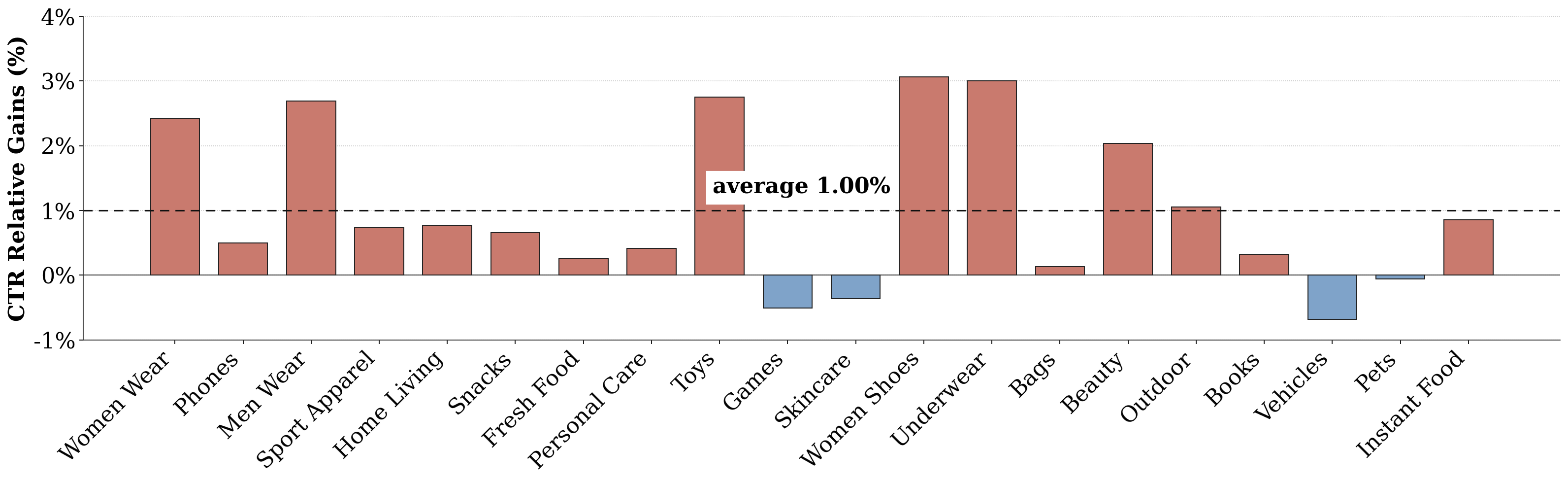}
\caption{Online CTR relative gains across the top $20$ industries; the dashed line marks the $1.00\%$ average.}
\label{fig:ctr_industries}
\end{figure}

\paragraph{Toward a Unified One-Model Recall Stage.}
Whether a single generative model can reach beyond all branches is the more ambitious question. A second deployment replaced both the inverted-index and the dense (vector) branches with OneRetrieval, under the same protocol as the first deployment: the offline unconstrained beam search, $20.0\%$ relative traffic (about $8.2\%$ absolute), and matched $7$-day AA and $11$-day AB windows. The second row of Table~\ref{tab:online} reports the result. The conversion metrics, Order and Buyer count, show no statistically significant change ($+0.255\%$ Order, $-0.028\%$ Buyer), while Item CTR rises significantly ($+0.821\%$). A single generative model thus replace nearly all retrieval branch, at no significant conversion loss and with a significant gain in click quality. Read together with the single-branch result, where replacing the inverted-index branch alone produced significant conversion gains, this two-tier evidence is directionally consistent with OneRetrieval subsuming the lexical and dense branch at no significant conversion cost. It points toward a recall stage consolidated into a single model, which would remove the hand-tuned cross-branch fusion and the cross-branch redundancy of the multi-branch architecture (Section~\ref{sec:intro}) while carrying the same-day editability of the inverted index into the generative regime.

\paragraph{Gains Are Broad-Based Across Industries.}
To examine where the click gains of the second deployment arise, we broke down its relative CTR gain across the top $20$ industries by query volume (Figure~\ref{fig:ctr_industries}). Sixteen of the twenty industries show positive gains, with an average relative improvement of $1.00\%$, so the benefit is broad-based rather than confined to a few categories. The largest gains concentrate in the apparel, footwear, and personal-care verticals, led by Women Shoes ($+3.05\%$), Underwear ($+3.00\%$), Toys ($+2.75\%$), Men Wear ($+2.70\%$), Women Wear ($+2.42\%$), and Beauty ($+2.05\%$). These are precisely the verticals in which queries carry rich attribute structure, such as brand, style, crowd, and color, and in which fine-grained product distinctions decide buyer intent; keyword-aligned encoding routes such queries to items through their attribute words and recovers matches that the lexical inverted-index branch misses, which is why the gain is largest here. The four industries with marginal negative effects, Games, Skincare, Vehicles, and Pets, are all bounded within $-0.7\%$ and are not statistically significant. Overall, replacing the lexical branch with keyword-aligned generative recall benefits nearly all industries, with the strongest effect in the intent-rich verticals where attribute matching matters most.

\section{Conclusion}
\label{sec:conclusion}
We presented OneRetrieval, a one-model generative retrieval framework for industrial e-commerce search that unifies the multi-branch recall stage while recovering, for the first time among generative methods to our knowledge, the real-time editability that has kept the inverted-index branch in production despite its below-average conversion. The framework rests on three coupled designs: Keyword-Aligned Encoding, which maps each identifier position to an interpretable key attribute word rather than a quantized embedding; an extensible codebook, which merges the key attribute categories information-theoretically, allocates capacity by attribute density, and reserves slots for parameter-free real-time intervention; and a four-stage supervised fine-tuning pipeline, which anchors the SID alphabet at Stage~0 and secures editability at Stage~2, so that the two goals are moved by near-disjoint stages.

Offline, OneRetrieval matches the strongest generative baseline at deep recall, surpasses the remaining dense and generative baselines by clear margins, exceeds closed-codebook encodings by more than an order of magnitude on intervention, and recovers most of the activation of the BM25 inverted index on identical injected terms; the modest precision gap at the shallow cut-offs is the deliberate price of this editability. Online, replacing the inverted-index branch alone yields statistically significant conversion gains, and extending the replacement to nearly the entire recall stage, leaves conversion unchanged while significantly lifting CTR, evidence that one generative model can begin to serve the multi-branch recall stage.

Future work includes extending the SID alphabet with multimodal signals to cover visually distinctive long-tail items, reinforcement learning over conversion signals to strengthen head queries, and learning to activate reserved slots from incoming traffic to automate trend response.


\begin{acks}
We thank the Kuaishou search platform team for infrastructure support and the operations team for participating in real-time intervention evaluations. During the preparation of this work the authors used Qwen3-14B to construct the primary-subject precedence judgments in the selection rules (Section~\ref{sec:method:codebook}) and Qwen3-7B to generate the out-of-vocabulary words used to build the offline intervention probe (Section~\ref{sec:exp:encoding}); all such outputs were verified by the authors, who take full responsibility for the content.
\end{acks}

\bibliographystyle{ACM-Reference-Format}
\bibliography{references}

\end{document}